\documentclass[runningheads]{llncs}
\usepackage{graphicx}
\usepackage{makecell}
\usepackage{float}
\usepackage{caption}
\usepackage{subcaption}
\usepackage{longtable}
\usepackage{tabularx}
\usepackage{tabu}
\usepackage{multirow}
\usepackage{tikz}
\usepackage{graphicx}
\usepackage{tikzscale}
\usepackage{pgfplots}
\usepackage[binary-units=true]{siunitx}
\usepackage{setspace}
\usepackage{booktabs}
\usepackage{siunitx}
\usepackage{adjustbox}
\usepackage{soul, color}
\usepackage{xcolor}
\usepackage[labelformat=simple]{subcaption}

\usepackage[most]{tcolorbox}
\newtcolorbox{highlighted}{colback=yellow,coltext=red,breakable}

\usepackage{algorithmic}
\usepackage{amsmath,amssymb,amsfonts}
\usepackage[ruled,linesnumbered]{algorithm2e}
\usepackage[font=small,labelfont=bf]{caption}
\SetKwFor{Foreach}{for each}{do}{endForEach}%
\SetKwFor{For}{for}{do}{endfor for}%
\SetKwIF{uIf}{ElseIf}{Else}{if}{then}{else if}{else}{endIf if}%
\SetKwFor{While}{while}{do}{endWhile while}%
\begin{document}
\title{Blockchain-based Access Control for Secure Smart Industry Management Systems}
%
%
\author{Aditya Pribadi Kalapaaking\inst{1}\orcidID{0000-0001-5796-8344} \and
Ibrahim Khalil\inst{1}\orcidID{0000-0001-5512-114X} \and
Mohammad Saidur Rahman\inst{1}\orcidID{0000-0002-4024-0725} \and Abdelaziz Bouras\inst{2}\orcidID{0000-0001-5765-1259}}
\authorrunning{A. P. Kalapaaking et al.}
%
\institute{School of Computing Technologies, RMIT University, Melbourne, Victoria, Australia\\ \email{aditya.pribadi.kalapaaking@student.rmit.edu.au} \and College of Engineering,
Qatar University, Doha,Qatar \email{abdelaziz.bouras@qu.edu.qa}}

%
\maketitle              
\begin{abstract}
Smart manufacturing systems involve a large number of interconnected devices resulting in massive data generation. Cloud computing technology has recently gained increasing attention in smart manufacturing systems for facilitating cost-effective service provisioning and massive data management. In a cloud-based manufacturing system,  ensuring authorized access to the data is crucial. A cloud platform is operated under a single authority. Hence, a cloud platform is prone to a single point of failure and vulnerable to adversaries. An internal or external adversary can easily modify users' access to allow unauthorized users to access the data. This paper proposes a role-based access control to prevent modification attacks by leveraging blockchain and smart contracts in a cloud-based smart manufacturing system. The role-based access control is developed to determine users' roles and rights in smart contracts. The smart contracts are then deployed to the private blockchain network. We evaluate our solution by utilizing Ethereum private blockchain network to deploy the smart contract. The experimental results demonstrate the feasibility and evaluation of the proposed framework's performance.

\end{abstract}
\section{Introduction}
The advancement of communication and Internet-of-Things (IoT) technologies has pushed the rapid development of smart industries or industry 4.0 to allow more efficient and customizable production and logistic operations. However, IoT devices collect a massive amount of data associated with their surroundings and influence cloud computing resources for storing and examining data to extract valuable insights.

Cloud technologies \cite{hayes2008cloud} is a superior technology that solves the problems in smart manufacturing systems. By offering the necessary platforms and infrastructure, cloud computing technology ensures efficient data management at a reasonable price. As a result, cloud service providers may manage data from smart manufacturing service providers effectively and affordably. 

However, cloud-based smart manufacturing systems bring up some security and trust issues. The centralized nature of cloud-based data storage enables a single authority to oversee the management of cloud-based data storage. As a result, the cloud service provider is susceptible to a single point of failure for the smart manufacturing services and stored transaction data. However, most cloud service providers employ advanced security measures to thwart outside cyberattacks. For example, an untrustworthy employee of the cloud service provider could alter or tamper with customer data or transactions. Therefore, internal cyberattacks against the cloud platform are not secure. 

Blockchain is a distributed system that links data structure for data storage, ensuring the data is resistant to modification. Initially, blockchain applications were limited to cryptocurrencies and financial transactions. The invention of smart contracts oversees the development of more diverse application scenarios such as healthcare \cite{rahman2020formalizing} and supply chains \cite{rahman2021framework}. Since blockchain is a decentralized system, it can solve a single point of failure from the cloud. To prevent resources stored in the cloud from being accessed or stolen by illegal users, access control is required for supplementary solutions. 

Therefore, a trustworthy smart manufacturing system is needed to guarantee the integrity of stored data and maintain the proper accessibility of the users' data in smart manufacturing management systems. The contributions of our work are summarized as follows:

\begin{itemize}
    \item provisioning of cloud-based search to aid in timely retrieval of end-user data
    \item blockchain-based storage to provide strong immutability for data provided by IoT devices
    \item access control powered by smart contracts to protect the ability of users to modify, view or delete the data
\end{itemize}

\section{Related Work}

Azaria et al. \cite{yue2016healthcare} proposed a decentralized management system, creating a prototype that showed an immutable database that provided access to their data via the facilities acting as miners in a blockchain network. These projects rely on proof-of-work (PoW) consensus processes, which demand a lot of processing power.

Although data retrieval is still computationally intensive, Wang et al. \cite{azaria2016medrec} developed the blockchain-based Data Gateway to encourage end-users to own, monitor, and exchange their data. Measa et al. \cite{maesa2017blockchain} also proposed using blockchain to publish and transfer resource usage rights regulations between users. However, they only used a hypothetical proof-of-concept Bitcoin implementation.

\cite{xia2017medshare} employed a blockchain-based framework to transmit data between institutions in the cloud, employing access protocols and smart contracts to track data movement and spot breaches. The lengthy response time of requests, which might take up to twenty minutes, is still a significant issue.

The issue of data protection on blockchain has been addressed. Zyskind et al. \cite{zyskind2015decentralizing} established a decentralized management system that uses a multi-party protocol for automatic access control to ensure users own their data.

The Ethereum ledger and attribute-based encryption technologies are combined in a decentralized storage architecture that Wang et al. 
\cite{wang2018blockchain} characterised as a blockchain-based data exchange architecture. Using the smart contracts on Ethereum,

A multi-authority attribute-based access management system was proposed by Guo et al. \cite{guo2019multi}

The FairAccess system was proposed by Ouaddah et al.\cite{ouaddah2016fairaccess}, and it uses a local ledger that is implemented on a Raspberry Pi computer to allow transactions to grant, obtain, delegate, and revoke access.

\begin{table}[tbh]
\caption{Overview of related work}
\label{rw}
\begin{tabularx}{1\textwidth} { 
  | >{\arraybackslash}X 
  | >{\arraybackslash}X 
  | >{\arraybackslash}X |
  }
    \hline
    \textbf{Model} & \textbf{Description} & \textbf{Remarks}\\
    \hline
    \cite{azaria2016medrec} & provide a decentralised management system for Health record, using permission management instead of access control & each node store specific information, however in the blockchain each node should have same record, permission management system consume huge computation resource\\
    \hline
    \cite{maesa2017blockchain} & provide a blockchain-based mechanism to published access rights & The access control still vulnerable to tampering attack, there are no experiment or evaluation set up for this concept in the paper\\
    \hline
    \cite{yue2016healthcare} & proposed a healthcare framework with utilising blockchain for the storage system & change the traditional database into the blockchain. However, the access control is placed on the end-user gateway which is not effective\\
    \hline
    \cite{xia2017medshare} & proposed a blockchain-based framework using a smart contract to authenticate the access for every query & in this paper, all the data still stored in a traditional database. They rely on the smart contract as an access control since in the architecture there is not any access control layer is used\\
    \hline
     
\end{tabularx}
\end{table}

Based on the current works, none of these earlier publications discuss the use of smart contracts with role-based access control systems, particularly with smart manufacturing based on the Internet of Things. In Table \ref{rw}, summarize the key point from our proposed method from the previous work.

\section{Proposed Framework}

This section presents our proposed blockchain-based access control. First, we present an overview of the system architecture. Next, we discuss the various components of our proposed framework in detail.

\subsection{Overview of the Proposed Framework}

\begin{figure}[tbh!]
\centering
\includegraphics[width=0.95\linewidth]{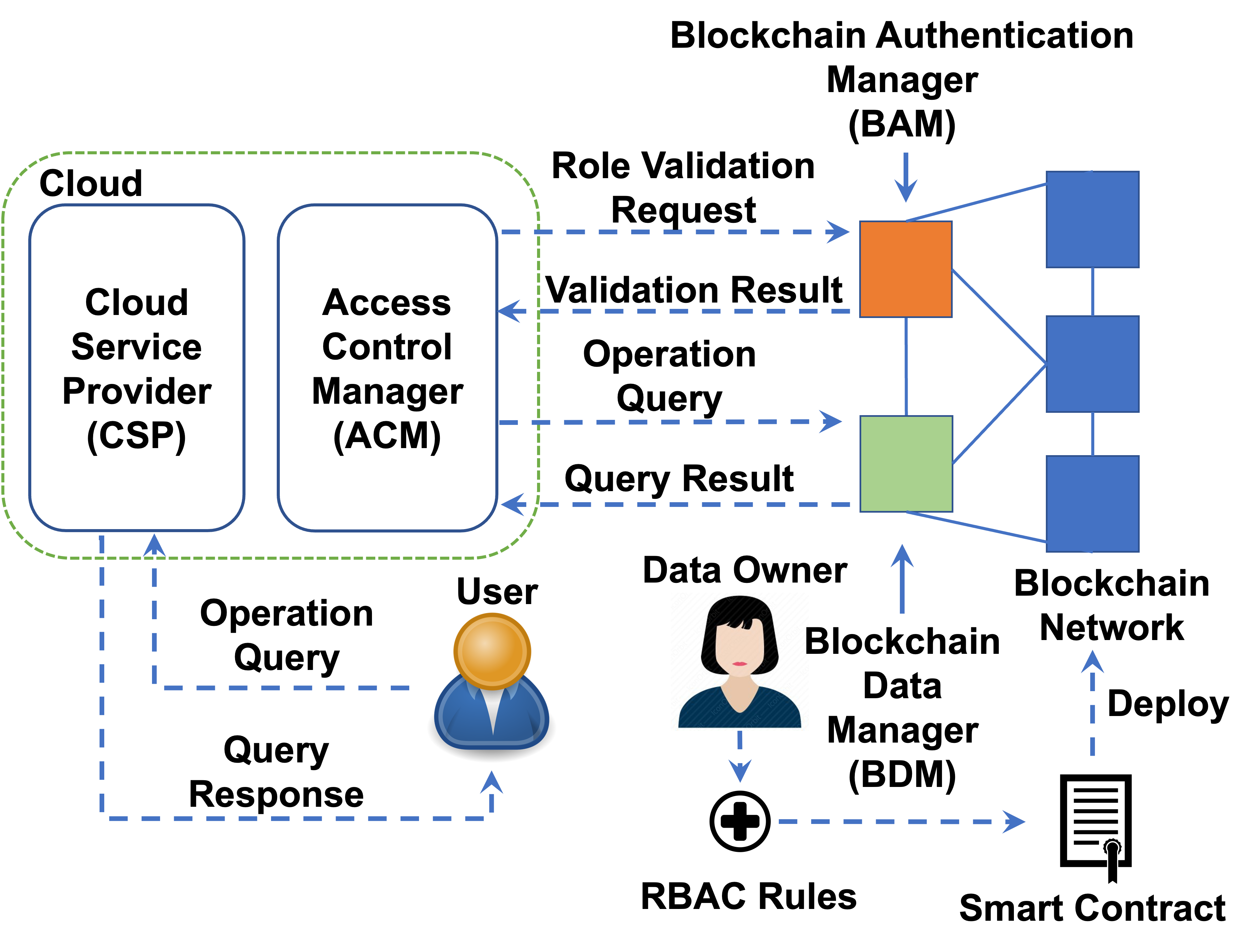}
\caption{Overview of the proposed framework}
\label{overview}
\end{figure}

The framework integrates blockchain and smart contract in cloud-based smart manufacturing management systems for tamper proof access control. There are two types of participants in the system: \textit{data owner} and \textit{users}. Data owners are the smart manufacturer (e.g., Administrator), whilst a user can be a engineer, product quality assurance, courier, supply-chain manager or even the customer itself. A large amount of data is generated by the various smart manufacturer sensors and devices, which authorised users gain access to for review or update.

Thus, it is paramount that user roles are verified given the sensitive nature of the data. However, role specifications themselves may be tampered and so, to ensure trustworthy definition and verification of roles, smart contracts are used.

Here, we integrate blockchain into the system, with smart contracts being deployed to all blockchain nodes. User authentication details (e.g., user id and roles) are stored within the blockchain to harden against modification attacks. Moreover, customers' transaction data is stored distributively amongst the blockchain nodes. Overview of the proposed framework is illustrated in Figure \ref{overview}.

To ease explanation in later sections, we break down the components of the proposed framework as follows:

\begin{itemize}
    \item \textbf{Cloud Service Provider (CSP)} acts as the interface between users and data, allowing direct communication and acting as an intermediary by answering user requests with results contained within. The actual query operation itself is performed by the ACM.
    
    \item \textbf{Access Control Manager (ACM)} only allows authorised access to data by users via requests. This involves user registration, role validation, and the retrieval of user rights. To ensure trustworthiness, it communicates with the blockchain network for user role validation. 
    
    \item \textbf{Blockchain Authentication Manager (BAM)} communicates with the ACM on behalf of the blockchain. It receives a user-role validation task as a transaction, communicating with other nodes to validate and retrieves the correct role with the help of smart contracts.
    
    \item \textbf{Blockchain Database Manager (BDM)} is another node in the blockchain network. Generally, it is responsible for executing transactions in the blockchain network, performing operations on blockchain data, and producing authenticated results.
\end{itemize}

\subsection{System Model}\label{sec:system_model}

Assume that a set of transaction records is stored as a blockchain in a blockchain network $BCN$. In the blockchain,  $U$ be the set of $n$ types of \textit{users} that is denoted as $U = \{u_1, u_2, \hdots, u_n\}$. $R$ be the set of $m$ \textit{roles} and denoted as $R = \{r_1, r_2, \hdots, r_m\}$. 

$R_{u_i}$ is the set of roles that is assigned to the user $u_i$. A role $r_i$ is associated with one or more \textit{rights}. The set $G$ of $k$ rights is denoted as $G = \{g_1, g_2, \hdots, g_k\}$. Let, $Att$ be the set of $l$ attributes in the transaction records that is denoted as: $Att = \{att_1, att_2, \hdots, att_l\}$. A right $g_i$ is the set of $l$ boolean values indicating the accessibility on $l$ different attributes of a transaction record. 

The formal definition of the Roles-based access control (RBAC) model is defined as follows:

\textbf{Definition 3.1} The RBAC model ($RBAC_M$) is a tuple: 
\begin{align}
    RBAC_M = <A_{U,R}, A_{R,G}, G>
\end{align}
where:
\begin{itemize}
    \item {$A_{U,R} = U \times R$, is the set of all possible \textit{role assignment relations} between users and roles. A user type $u_i \in U$ can be assigned to roles $\{r^{'}|r^{'} \subseteq R\}$, then the role assignment can be denoted as $A_{u_{i},r^{'}} \subseteq A_{U,R}$.}
    \item {$A_{R,G} = R \times G$, is the set of all possible \textit{right association relations} between roles and rights. A role $r_i \in R$ is associated with rights $\{g^{'}|g^{'} \subseteq G\}$, then the right association can be denoted as $A_{r_{i},g^{'}} \subseteq A_{R,G}$.}
    \item {$g_i \in G$ can be denoted as $g_i = \{b_1, b_2, \hdots, b_l\}$ where $b_j$ is the accessibility indicator on the $j$th attribute ($1 \leq j \leq l$) and $b_j = true$ or $false$. The value $b_j = true$ indicates that the $j$th attribute $att_j \in Att$ is accessible and $b_j = false$ indicates otherwise.}
\end{itemize}

The accessibility of a particular type of user is verified based on the $RBAC_M$. To obtain the accessibility, a user needs to send a data access request. The \textit{data access request} can be formally defined as follows: 

\textbf{Definition 3.2} A data access request is a function $req(p,param)$, where $p$ is the unique user ID with a user type $p.type$, and $param$ is the list of query parameters containing $q$ attribute ($att$) and value ($val$) pairs. $param$ can be represented as: $param = \{(att_1,val_1), \\ (att_2,val_2), \hdots, (att_q,val_q)\}$.  

\textbf{Definition 3.3} An \textit{Accessibility Rule} is a set $AR$ of semantics that must be satisfied to verify the accessibility of the user $p$ while sending a request $req(p,param)$. $AR$ comprises of the following semantics:
\begin{align}
    p.type \subseteq U,
\end{align}
\begin{align}
    p.role \subseteq A_{U,R} \ \text{and} \  p.role = A_{u_{p},r_{p}}. 
\end{align}
\begin{align}
    p.rights \subseteq A_{R,G} \ \text{and} \ p.rights = A_{r_{i},g^{'}}.
\end{align}
\begin{align}
    \forall{att_i \in param.Att^{'}}, param.Att^{'} \subseteq Att \land att_i \in g^{'}.
\end{align}

The $RBAC_M$ model is defined in a \textit{smart contract} (SC) to ensure the trustworthy validation of accessibility based on the semantics $AR$. The formal definition of the smart contract can be provided as below:

\textbf{Definition 3.4} A smart contract ($SC$) is a tuple: 
\begin{align}
    SC = <Op, AR>,
\end{align}
where $Op$ is set of operations in $SC$ and $AR$ is the set of accessibility rules.

\subsection{Role-based Access Control using Smart Contract} \label{sec:sc_rbac}

This section describes the proposed role-based access control using smart contracts. The proposed role-based access control mechanism involves several steps. Each step is described below:

\subsubsection{Step-1: \textit{Initialization}}
The data owner, the access control layer ($ACM$), the blockchain authentication manager ($BAM$), and the blockchain data manager ($BDM$) all generate keys as part of the initialization process. Assume that $KeyGen()$ is a key generation algorithm based on public-key cryptography that generates a public-key ($PK$) and private-key ($PR$) pair. Using $KeyGen()$, the data owner creates a key pair made up of a public key ($PK_{DO}$) and a private key ($PR_{DO}$). The data owner will use this key pair to deploy the smart contract and carry out its activities. The data owner preserves $PR_{DO}$ as a secret and shares $PK_{DO}$ with the cloud's $ACM$ and the blockchain network's $BAM$. Using $KeyGen()$, $ACM$ also creates a key pair, including a public key ($PK_{ACM}$) and a private key ($PR_{ACM}$). The key-pairs for $BAM$ and $BDM$ are generated similarly as follows: $PK_{BAM}$, $PR_{BAM}$, and $PK_{BDM}$, $PR_{BDM}$.

\subsubsection{Step-2: \textit{Generation and deployment of smart contracts}}
\begin{figure}[tbh!]
\centering
\includegraphics[width=0.95\linewidth]{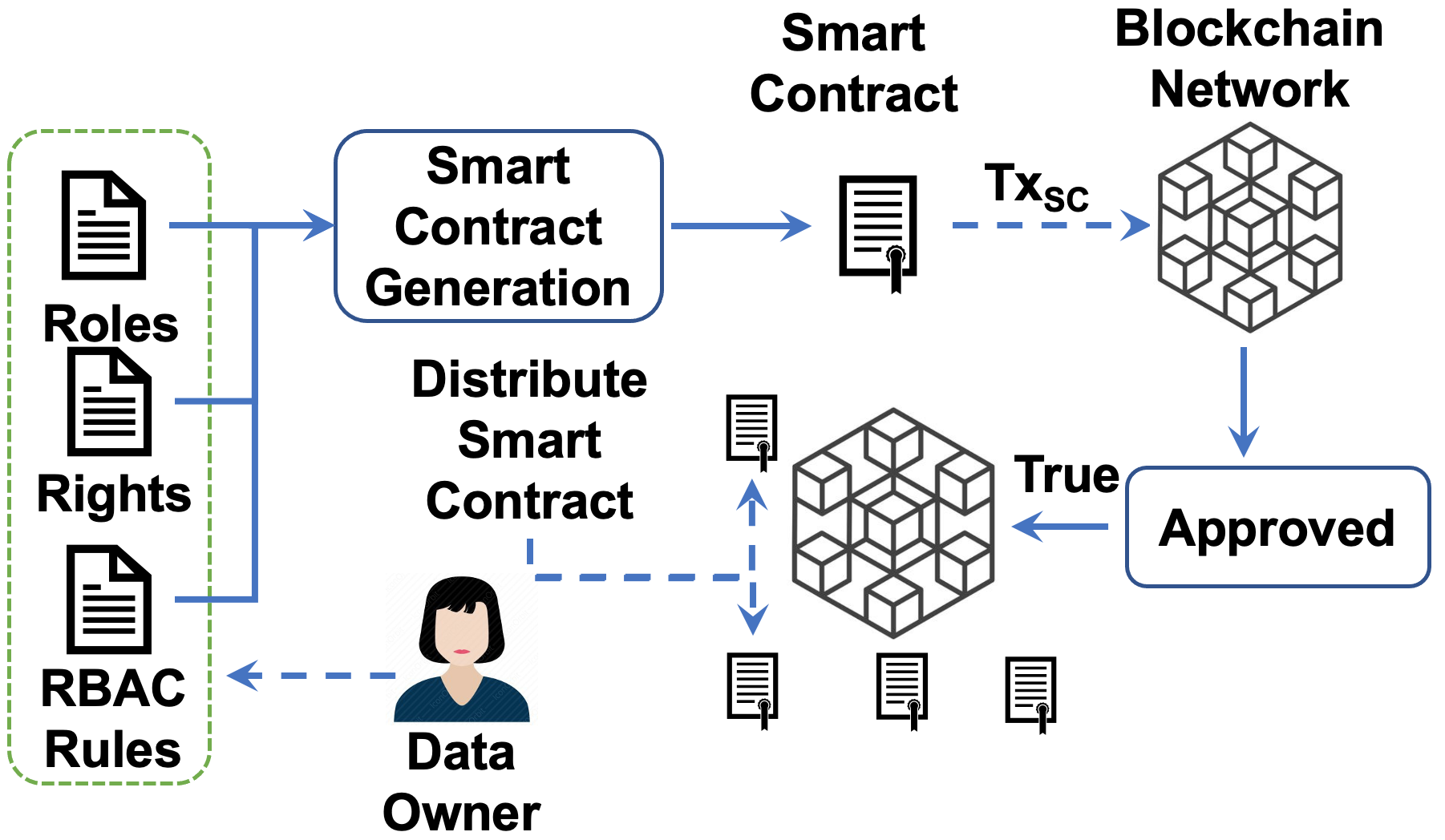}
\caption{Generation and deployment of Smart Contract}
\label{fig:sc_gen_deploy}
\end{figure}

In this stage, the data owner establishes roles for various user types and sets the access control rules ($AR$) depending on those user types. The data owner then uses $AR$ as described in Definition 3.4 to create a smart contract (SC). In order to deploy $SC$ in the blockchain network, the data owner sends a transaction to the BAM called $Tx_{SC}$. Formally, $Tx_{SC}$ can be written as follows:
\begin{align}
    Tx_{SC} = \{ID_{DO}, ID_{BAM}, cost_{SC}, Sign(SC,PR_{DO})\},
\end{align}
where:
\begin{itemize}
    \item {$ID_{BAM}$ is the unique ID of the blockchain authentication manager ($BAM$).}
    \item {$ID_{DO}$ is the unique ID of the data owner representing the transaction generator.}
    \item {$cost_{SC}$ is the price of running the transaction $TX_{SC}$.}
    \item {$Sign(SC,PR_{DO})$  is the signed smart contract $SC$ that is constructed using a digital signature technique with the data owner's private key $PR_{DO}$. }
\end{itemize}
A smart contract scripting language for the blockchain platform can be used to create $SC$. The blockchain platform for this study is Ethereum, and $SC$ is created using the Solidity programming language. Section \ref{sec:setup} discusses the implementation specifics. Figure \ref{fig:sc_gen_deploy} shows the $SC$ creation and deployment operations.

\subsubsection{Step-3: \textit{User registration}}
By giving each user one or more roles in this stage, the data owner creates new users and registers them. When a user is created, roles are assigned, and the user roles are stored on the blockchain. The data owner issues a blockchain transaction called $Tx_{UR}$. The Blockchain Data Manager ($BDM$) receives $Tx_{UR}$ from the data owner, which is forwarded to the blockchain network for inclusion in the blockchain. The following is a formal representation of $Tx_{UR}$:
\begin{align}
    Tx_{UR} = \{ID_{DO}, ID_{BDM}, cost_{UR}, Sign(U_{pro}, PR_{DO})\},
\end{align}
where:
\begin{itemize}
\item {$ID_{BDM}$ is the unique ID of the blockchain data manager ($BDM$).}
    \item {$ID_{DO}$ is the unique ID of the data owner representing the transaction generator.}
    \item {$cost_{UR}$ is the price of running the transaction $TX_{UR}$.}
    \item {$Sign(U_{pro},PR_{DO})$ is the signed user data $U_{pro}$ that is produced using a digital signature schemes with data owner's private-key $PR_{DO}$. $U_{pro}$ is the digitally signed user data $U_{pro}$ that was created using the private key of the data owner $PR_{DO}$. The list of user data known as $U_{pro}$ is denoted as $U_{pro} = \{p, R^{'}_{p}\}$, where $p$ is the user's unique ID and $R^{'}_{p}$ is the set of user roles such that $R^{'}_{p} \subseteq R$.}
\end{itemize}
The only transactions $Tx_{UR}$ that the data owner can create are those for adding users and assigning roles. Before storing $ Tx_{UR}$ into the blockchain, the network verifies it. As a result, malicious users cannot create fake user roles or alter those that already exist.    

\subsubsection{Step-4: \textit{User role validation and granting access}}
In this stage, a user role is verified, and access is granted to a system-authorized user. After signing up for the system, a user can send a data access request to the cloud service provider ($CSP$) using the syntax $req(p, param)$. The access control layer ($ACM$) receives $req(p, param)$ from the cloud service provider. The user role validation transaction $Tx_{V}$ is then created by $ACM$ and sent to $BAM$ for user role validation. The following is a representation of $Tx_{V}$:
\begin{align}
\begin{split}
    Tx_{V} = \{ID_{ACM}, ID_{BAM}, cost_{V},\\ID_{SC}, Sign(p.role, PR_{ACM})\},
\end{split}
\end{align}
where:
\begin{itemize}
    \item {$ID_{ACM}$ is the unique ID of $ACM$ denoting the transaction generator.}
    \item {$ID_{BAM}$ is the unique ID of the blockchain authentication manager ($BAM$).}
    \item {$cost_{V}$ is the cost of executing the transaction $TX_{V}$.}
    \item {$ID_{SC}$ is the smart contract's unique ID.}
    \item {$Sign(p.role,PR_{ACM})$ is the signed user role that is produced using a digital signature schemes with ACM's private-key $PR_{ACM}$.}
\end{itemize}
In the blockchain network, $BAM$ propagates $Tx_{V}$ to verify user roles. Using the smart contract $SC$, the user role is verified. The matching rights ($p.rights$) of the user are then returned to $BAM$, signed by $BAM$, and sent to $ACM$. $Sign(p.rights, PR_{BAM})$ can be used to represent the signed rights. $ACM$ then gives the user $p$ the rights. The user role validation procedure based on smart contracts is represented by the Algorithm \ref{alg:validation}.

\begin{algorithm}
\caption{Smart Contract based user role validation process}
\label{alg:validation}
\KwIn
{
    \begin{minipage}[t]{10cm}%
     \strut
      $p.type$, type of current user $p$ 
     \strut
  \end{minipage}%
}
\KwOut
{
    \begin{minipage}[t]{10cm}%
     \strut
      $p.rights$, rights of current user $p$ 
     \strut
  \end{minipage}%
}

\While{CSP}
{
    \eIf{ $p.type \subseteq U$}
    {
        \eIf{ $p.role \subseteq A_{U,R} \wedge p.role = A_{u_{p},r_{p}}$ }
        {
            $Send Tx_{V} = {ID_{ACM}, ID_{BAM}, cost_{V}, ID_{SC}, Sign(p.role, PR_{ACM})} to SC$
            \eIf{$Tx_{V}$ is valid}
            {
             return $p.rights$
             return $p.rights = \text{NULL}$
                 \If{$p.rights \neq NULL$}
                 {
                 $ACM$ grants access to user $p$ with $p.rights$
                  }
            }{$p.rights \neq valid$}
        }{Invalid User}
    }{Invalid User}
}
\end{algorithm}

\subsubsection{Step-5: \textit{Accessibility based Operation}}
\begin{figure}[tbh!]
\centering
\includegraphics[width=1\linewidth]{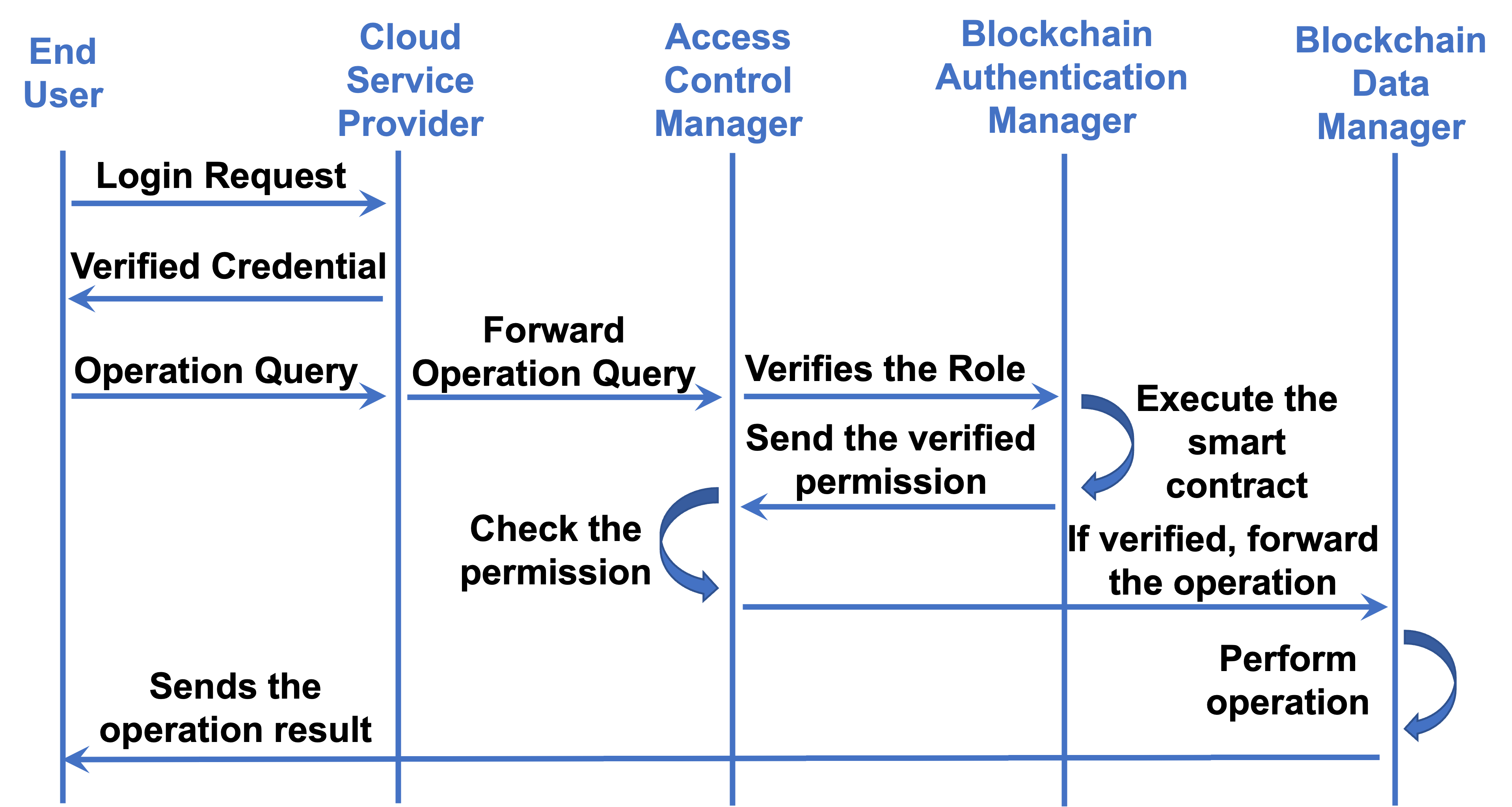}
\caption{Accessibility based operation}
\label{commenduser}
\end{figure}

The user does the action in this stage in accordance with the responsibilities and rights granted to the user. The sequence diagram for the entire accessibility-based operation is shown in Figure \ref{commenduser}. An end user is initially authenticated by the CSP. The end user then issues an ACM inquiry request. The ACM retrieves the rights for the user and confirms the roles with BAM. After then, the ACM sends the query request to the BDM for processing. For the request, the BDM creates a query result and transmits it to the CSP. The end-user receives the result from CSP.

\section{Experimental Results}\label{chap:experiment}
In this section, we show several experiments conducted to evaluate the performance of our proposed framework and discuss the results.

\subsection{Experimental Setup}\label{sec:setup}
In our experiments, we run the experiments with the AWS EC2 cloud. We use \textit{T2.XLarge} instance and it has 4 vCPU and 1 GB of RAM that simulate medium size smart manufacturer server.

The implementation of the prototype comprises of the \textit{core framework} and the \textit{blockchain}. The core framework is developed as a Java server-side application, which is then interfaced with an Ethereum blockchain emulated in Ganache \cite{lee2019testing}, which includes ten accounts as default with 100 ethers. Accounts and held Ethereum sums may be changed as needed.

Each account can send and receive Ethereum transfers, or engage in smart contract activities. By forming a block for each operation, the Ganache blockchain also provides miner consent. Therefore, it is not required to wait for the transfers to be approved in the virtual environment \cite{karatas2018developing}. We develop and communicate smart contracts using both Solidity \cite{mukhopadhyay2018ethereum} and the Truffle \cite{wimmer2012truffle} framework. NodeJS \cite{tilkov2010node} is used to communicate between servers and Ethereum nodes. To enable communication between the Java-based core framework and the Ethereum blockchain environment, Web3j \cite{labs} is used.

\subsection{Results and Performance Evaluation}

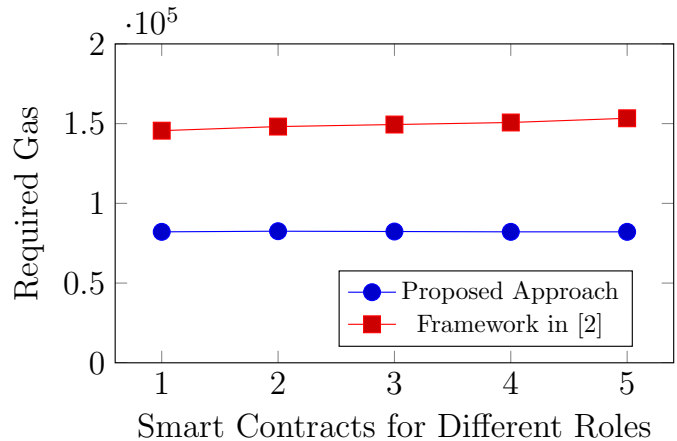
\begin{figure}[tbh]
\centering
    \resizebox{0.75\columnwidth}{!}{
    \begin{tikzpicture}
    \pgfplotsset{
    scale only axis,
    compat=1.5,
    width=7cm,
    height=4cm,
    }
    \begin{axis}[
    xlabel={Smart Contracts for Different Roles},
    ylabel={Required Gas},
    symbolic x coords = {1, 2, 3, 4, 5,6,7,8,9,10,11,12,13,14,15,16,17,18,19,20},
    xtick=data,
    ymax=200000,
    ymin=0,
    legend pos=north west,
    grid style=dashed,
    legend style={nodes={scale=0.9, transform shape},at={(0.4,0.05)},anchor=south west},
    label style={font=\large},
    tick label style={font=\large}
]
\addplot+[mark size=3pt]
    coordinates {
    (1,82129)
    (2,82529)
    (3,82329)
    (4,82129)
    (5,82129)
    };
\addplot+[mark size=3pt]
    coordinates {
    (1,145590)
    (2,148152)
    (3,149432)
    (4,150712)
    (5,153351)
    };
    \legend{Proposed Approach, Framework in \cite{cruz2018rbac}}
    \end{axis}
    \end{tikzpicture}
    }
    \caption{Required Gas to generate smart contracts}
    \label{expgas}
\end{figure}

To begin, we examine the cost to generate the smart contracts. Cost here is computed in terms of \textit{Gas}, the unit used in the Ethereum network. For comparison, the deployment cost of the systems proposed by Cruz \textit{et al.} in \cite{cruz2018rbac} is shown in Figure \ref{expgas} alongside the costs incurred by the proposed approach. As seen, results indicate that our generation costs consistently track 50\% lower across a similar number of roles. This bodes well for the cost-effectiveness of our approach.

\begin{figure}[b]
\centering
    \resizebox{0.625\columnwidth}{!}{
    \begin{tikzpicture}
    \pgfplotsset{
    scale only axis,
    compat=1.5,
    width=7cm,
    height=3cm,
    }
    \begin{axis}[
    xlabel={Simultaneous Requests},
    ylabel={Execution Time (ms)},
    symbolic x coords = {100,200,300,400,500},
    xtick=data,
    ymax=100000,
    ymin=10000,
    legend pos=north west,
    grid style=dashed,
    legend style={nodes={scale=1.0, transform shape}},
    label style={font=\large},
    tick label style={font=\large}
    ]
    \addplot+[mark size=3pt]
    coordinates {
    (100,28931)
    (200,59625)
    (300,86998)
    };
    \legend{}
    \end{axis}
\end{tikzpicture}
    }
 \caption{Execution time for query requests}
 \label{expsearch}
\end{figure}
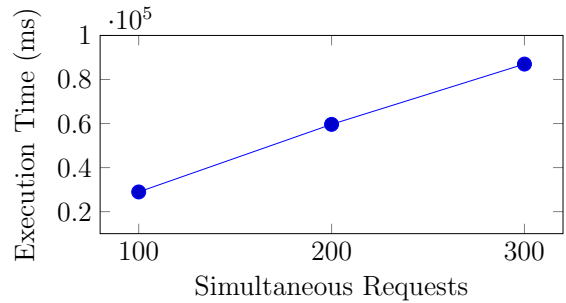

Next, we examine at the time cost to produce the query results in Figure \ref{expsearch} where several concurrent query operations are performed and then transmitted to the ACM to be processed. The goal is to imitate concurrent consumers querying the blockchain for particular sets of data. All queries are forwarded via the ACM to the BDM, which controls network-wide query activities. The timings displayed reflect projected peaks and troughs within operational times for different simultaneous query requests of 100 to 300. According to the results, execution times rise linearly as the number of requests increases. Even at the upper end, we saw responses in 86 seconds for 300 simultaneous requests, which would serve all users in a typical mid-size smart manufacturing. Please note that when the system is implemented on more potent machines, these timings will dramatically improve.

\pgfplotsset{
    my axis style/.style={
        axis line style={black!70},
        every axis label={font=\small},
        xlabel style={yshift=0.5em},
        ylabel style={yshift=-0.5em},
        tick label style={font=\small},
        label style={font=\small},
        legend style={font=\small},
    }
}
\begin{figure}[th!]
\centering
    \resizebox{1.0\columnwidth}{!}
    {
    \begin{tikzpicture}
    \pgfplotsset{
    scale only axis,
    compat=1.5,
    xmin=1, xmax=20,
    my axis style,
    width=7cm,
    height=3.5cm,
}
\begin{axis}[
  axis y line*=left,
  ymin=0, ymax=20,
  xlabel=User Rights per Role,
  ylabel=Verification Time (ms),
  y axis line style=red,
  y tick label style=red
]
\addplot[smooth,mark=x,red]
  coordinates{
    (20,15)
    (19,13)
    (18,11)
    (17,11)
    (16,9)
    (15,9)
    (14,7)
    (13,7)
    (12,9)
    (11,9)
    (10,7)
    (9,8)
    (8,8)
    (7,7)
    (6,7)
    (5,7)
    (4,6)
    (3,6)
    (2,6)
    (1,5)
};\label{exp8}
\end{axis}

\begin{axis}[
  axis y line*=right,
  axis x line=none,
  ymin=100, ymax=140,
  ylabel=Deployment Time (ms),
  y axis line style=blue,
  y tick label style=blue,
  legend style={nodes={scale=1.0, transform shape},at={(0.05,0.9)},anchor=north west},
]
\addlegendimage{/pgfplots/refstyle=exp8}\addlegendentry{verify}
\addplot[smooth,mark=*,blue]
  coordinates{
    (20,130)
    (19,125)
    (18,123)
    (17,122)
    (16,120)
    (15,120)
    (14,121)
    (13,122)
    (12,121)
    (11,121)
    (10,118)
    (9,118)
    (8,120)
    (7,117)
    (6,119)
    (5,118)
    (4,116)
    (3,117)
    (2,117)
    (1,116)
};
\label{exp2}
\addlegendentry{deploy}
\end{axis}
\end{tikzpicture}
}
\caption{Deployment and verification times of specific usage rights within user roles}
\label{exp8Nexp2}
\end{figure}
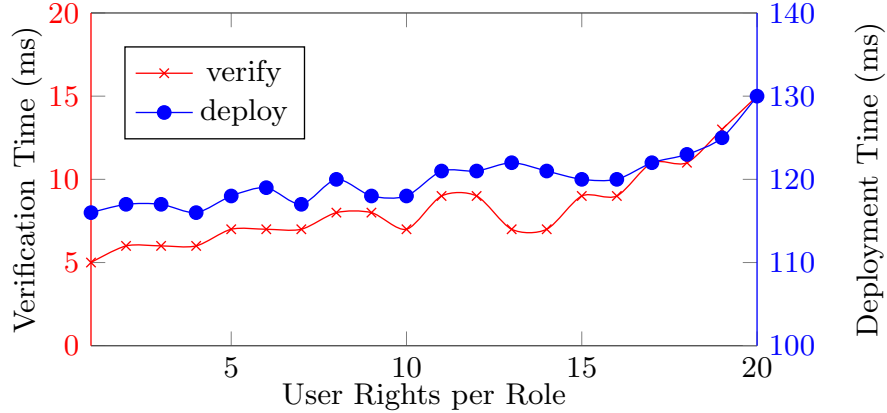

Figure \ref{exp8Nexp2} shows the execution times for the deployment (\ref{exp2}) and verification (\ref{exp8}) of smart contracts in our framework. In this experiment, blockchain-deployed smart contracts are made for various permissions within certain positions. We also confirm their rights in order to be thorough. Giving the admin permission to complete the order after carefully reviewing the quality and quantity is one example. Therefore, the client can inform the manufacturer if there is an issue with the products (such as a wrong amount or a damaged item).

The deployment phase, which demands around 115 to 130 ms over the 20 usage rights we defined, is more time-consuming, as was to be expected. With a speed range of 5 to 15 ms, the verification phase is quicker. Both phases have a slight rising trend, with the time taken growing as more rights are added. With the manager, accountant, technician, and administrative personnel all having distinct and purposeful usage rights within their responsibilities, it is envisaged that the chosen number of 20 will more than satisfy standard smart manufacturer requirements.

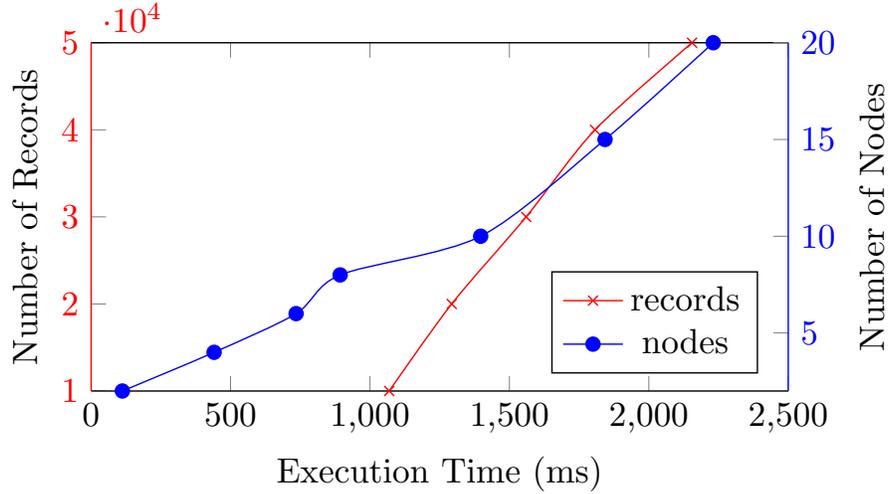
\begin{figure}[t]
\centering
    \resizebox{1.0\columnwidth}{!}{
    \begin{tikzpicture}
    \pgfplotsset{
    compat=1.5,
    scale only axis,
    xmin=0, xmax=2500,
    width=7cm,
    height=3.5cm,
}
\begin{axis}[
  axis y line*=left,
  ymin=10000, ymax=50000,
  xlabel=Execution Time (ms),
  ylabel=Number of Records,
  y axis line style=red,
  y tick label style=red,
]
\addplot[smooth,mark=x,red]
  coordinates{
    (1069,10000)
    (1293,20000)
    (1560,30000)
    (1807,40000)
    (2155,50000)
};
\label{exp1a}
\end{axis}

\begin{axis}[
  axis y line*=right,
  ymin=2, ymax=20,
  axis x line=none,
  ylabel=Number of Nodes,
  y axis line style=blue,
  y tick label style=blue,
  legend style={nodes={scale=1.0, transform shape},at={(0.66,0.05)},anchor=south west}
]
\addlegendimage{/pgfplots/refstyle=exp1a}\addlegendentry{records}
\addplot[smooth,mark=*,blue]
  coordinates{
    (112,2)
    (441,4)
    (735,6)
    (893,8)
    (1397,10)
    (1843,15)
    (2231,20)
};
\label{exp1b}
\addlegendentry{nodes}
\end{axis}
\end{tikzpicture}
    }
    \caption{Generation times across number of records and nodes}
    \label{exp1aNexp1b}
\end{figure}

In Figure \ref{exp1aNexp1b} we analyze the time needed to generate the blockchain for the smart manufacturer data across various amounts of records (\ref{exp1a}) and nodes (\ref{exp1b}). The objective is to monitor performance as we scale up the number of participating nodes and records in the network. We scale the number of records from 10,000 to 50,000 after limiting the number of blockchain nodes to 4. As a result, we observe a rising linear trend, with 10K records starting at 1.069 seconds and 50K records ending at 2.155 seconds. This indicates the potential of incorporating a substantial amount of manufacturer-provided IoT data into the system, such as access control and consumer transactions. The processing time will be shortened by using a full server with greater power.

Finally, in the same Figure \ref{exp1aNexp1b}, we observe the impact of including more blockchain nodes in the system. In this instance, we set the record count at 10K and test against 2 to 20 nodes. Once more, we observe a linear growth from two nodes at 112 ms to twenty nodes at 2231 ms. These results indicate the viability of the strategy if scaling up to a more extensive blockchain is necessary, keeping in mind the hardware limits once more.

\subsection{Discussion}\label{chap:discussion}

The proposed framework integrates blockchain and smart contract technology. This ameliorates some of the known issues with centralized cloud platforms as we seek to decentralize important access-control mechanisms and thus harden them against attacks. The immutability afforded by the blockchain is a crucial pillar of this framework, with malicious actors facing an uphill task if they wish to tamper with actual data. Further, as the roles and rights of system users are defined in smart contracts, which in turn are also replicated to all nodes in the blockchain, attacks such as privilege escalations or false authorizations are minimized.

Each data request is submitted as a blockchain transaction in the suggested architecture. These transactions comprise the creation of smart contracts ($Tx_{SC}$), registration of users ($Tx_{UR}$), and validation of user roles ($Tx_{V}$). The associated transaction generator uses public-key cryptography to sign each of the transactions above digitally. The signature will be secure if the correct key settings are applied. Using the aggregated key, users can confirm the search result. Because of this, even a 51 percent attack cannot change the query result. Hence, the proposed framework guarantees the verifiability of the query result. 

Results presented in Section \ref{chap:experiment}, highlight the efficacy and performance of the proposed blockchain-enabled role-based access control (RBAC). The experiments show that execution times for most operations (i.e., generation and verification of user roles) follow a linear trend without spikes as we increase the number of records and nodes within the network. With the limited source in mind, we postulate that a larger variant of the proposed approach would perform better in an environment with more powerful machines. It is reasonable to expect that significant smart manufacturing would be able to accommodate the computational requirements.

\section{Conclusion}
In this paper, a blockchain-based access control framework is proposed to ensure the integrity of the data and transaction within the context of a smart manufacturing. First, a decentralized data storage model is introduced that stores the transactions records in the blockchain. The blockchain of records are replicated across multiple nodes to ensure integrity and to protect against tampering. Second, a smart contract-based access control mechanism is proposed to define the roles of different system users. Different roles and their corresponding rights can be created and stored in multiple smart contracts to be deployed in the blockchain network. The smart contracts are replicated amongst nodes in the network, with user role creation and validation tasks generated only via valid blockchain transactions. Accordingly, false user roles cannot be created and none of the existing user roles can be modified by an attacker. As seen in the experimental results, the proposed role-based access control (RBAC) using smart contracts is cost-effective. Moreover, execution times for smart contract generation and verification tasks showed linear characteristics, which points both to the efficiency and scalability of the approach. The hierarchy of roles necessary for the beneficial role and proper management is not taken into account by the current approach. Our future goal is to include a hierarchical model for roles and rights management.

\section*{Acknowledgement}
This work is part of the NPRP11S-1227-170135 project.
The authors would like to express their gratitude to the QNRF (Qatar Foundation) for its support and funding
for the project activities.

\bibliographystyle{splncs04}
\bibliography{References}

\begin{thebibliography}{10}
\providecommand{\url}[1]{\texttt{#1}}
\providecommand{\urlprefix}{URL }
\providecommand{\doi}[1]{https://doi.org/#1}

\bibitem{azaria2016medrec}
Azaria, A., Ekblaw, A., Vieira, T., Lippman, A.: Medrec: Using blockchain for
  medical data access and permission management. In: 2016 2nd International
  Conference on Open and Big Data (OBD). pp. 25--30. IEEE (2016)

\bibitem{cruz2018rbac}
Cruz, J.P., Kaji, Y., Yanai, N.: Rbac-sc: Role-based access control using smart
  contract. Ieee Access  \textbf{6},  12240--12251 (2018)

\bibitem{guo2019multi}
Guo, H., Meamari, E., Shen, C.C.: Multi-authority attribute-based access
  control with smart contract. In: Proceedings of the 2019 International
  Conference on Blockchain Technology. pp. 6--11 (2019)

\bibitem{hayes2008cloud}
Hayes, B.: Cloud computing (2008)

\bibitem{karatas2018developing}
Karatas, E.: Developing ethereum blockchain-based document verification smart
  contract for moodle learning management system. Online Submission
  \textbf{11}(4),  399--406 (2018)

\bibitem{labs}
Labs, W.: Web3j¶, \url{https://docs.web3j.io/}

\bibitem{lee2019testing}
Lee, W.M.: Testing smart contracts using ganache. In: Beginning Ethereum Smart
  Contracts Programming, pp. 147--167. Springer (2019)

\bibitem{maesa2017blockchain}
Maesa, D.D.F., Mori, P., Ricci, L.: Blockchain based access control. In: IFIP
  international conference on distributed applications and interoperable
  systems. pp. 206--220. Springer (2017)

\bibitem{mukhopadhyay2018ethereum}
Mukhopadhyay, M.: Ethereum Smart Contract Development: Build blockchain-based
  decentralized applications using solidity. Packt Publishing Ltd (2018)

\bibitem{ouaddah2016fairaccess}
Ouaddah, A., Abou~Elkalam, A., Ait~Ouahman, A.: Fairaccess: a new
  blockchain-based access control framework for the internet of things.
  Security and Communication Networks  \textbf{9}(18),  5943--5964 (2016)

\bibitem{rahman2020formalizing}
Rahman, M.S., Khalil, I., Bouras, A.: Formalizing dynamic behaviors of smart
  contract workflow in smart healthcare supply chain. In: International
  Conference on Security and Privacy in Communication Systems. pp. 391--402.
  Springer (2020)

\bibitem{rahman2021framework}
Rahman, M.S., Khalil, I., Bouras, A.: A framework for modelling blockchain
  based supply chain management system to ensure soundness of smart contract
  workflow. In: HICSS. pp. 1--10 (2021)

\bibitem{tilkov2010node}
Tilkov, S., Vinoski, S.: Node. js: Using javascript to build high-performance
  network programs. IEEE Internet Computing  \textbf{14}(6),  80--83 (2010)

\bibitem{wang2018blockchain}
Wang, S., Zhang, Y., Zhang, Y.: A blockchain-based framework for data sharing
  with fine-grained access control in decentralized storage systems. Ieee
  Access  \textbf{6},  38437--38450 (2018)

\bibitem{wimmer2012truffle}
Wimmer, C., W{\"u}rthinger, T.: Truffle: a self-optimizing runtime system. In:
  Proceedings of the 3rd annual conference on Systems, programming, and
  applications: software for humanity. pp. 13--14 (2012)

\bibitem{xia2017medshare}
Xia, Q., Sifah, E.B., Asamoah, K.O., Gao, J., Du, X., Guizani, M.: Medshare:
  Trust-less medical data sharing among cloud service providers via blockchain.
  IEEE Access  \textbf{5},  14757--14767 (2017)

\bibitem{yue2016healthcare}
Yue, X., Wang, H., Jin, D., Li, M., Jiang, W.: Healthcare data gateways: found
  healthcare intelligence on blockchain with novel privacy risk control.
  Journal of medical systems  \textbf{40}(10), ~218 (2016)

\bibitem{zyskind2015decentralizing}
Zyskind, G., Nathan, O., et~al.: Decentralizing privacy: Using blockchain to
  protect personal data. In: 2015 IEEE Security and Privacy Workshops. pp.
  180--184. IEEE (2015)

\end{thebibliography}
\end{document}